\documentclass[12pt]{iopart}
\usepackage{epsfig}
\begin{document}
\title{ \bf  The mobility of dual vortices in honeycomb, square, triangular, Kagome and
             dice lattices  }
\author{ \bf  Longhua Jiang and Jinwu Ye  }
\address{  Physics Department, The Pennsylvania State University, University Park, PA, 16802 }
\date{\today}

\begin{abstract}
    It was known that by a duality transformation, interacting bosons at filling factor $ f=p/q $ hopping on a lattice
    can be mapped to interacting vortices hopping on the dual lattice subject to a fluctuating
    {\em dual} " magnetic field" whose average strength through a dual plaquette is equal to the boson density $ f=p/q $.
    So the kinetic term of the vortices is the same as the Hofstadter problem
    of electrons moving in a lattice in the presence of  $
    f=p/q $ flux per plaquette. Motivated by this mapping, we study the Hofstadter
    bands of vortices hopping in the presence of magnetic flux
    $ f=p/q $ per plaquette on 5 most common bipartite and frustrated lattices namely square, honeycomb, triangular,
    dice and Kagome lattices. We count the total number of bands, determine the number of minima
    and their locations in the lowest band. We also numerically calculate
    the bandwidths of the lowest Hofstadter bands in these
    lattices that directly measure the mobility of the dual
    vortices. The less mobil the dual vortices are, the more likely in a superfluid state the bosons are.
    We find that except the Kagome lattice at odd $ q $,  they  all satisfy the exponential decay
    law $ W = A e^{-cq} $ even at the smallest $ q $. At given $ q $, the bandwidth $ W $ decreases in the order of
    Triangle, Square and Honeycomb lattice. This indicates that the domain of
    the superfluid state of the original bosons increases in the order of the corresponding direct
    lattices: Honeycome, Square and Triangular.
    When $ q=2 $, we find that the the lowest Hofstadter band
    is completely flat for both Kagome and dice lattices. There is a gap on Kagome lattice, but no gap on dice
    lattice. This indicates that the  boson ground state  at half filling with nearest neighbor hopping on Kagome lattice
    is always a superfluid state. The superfluid state remains
    stable slightly away from the half filling. Our results show
    that the behaviours of bosons  at or near half filling  on Kagome lattice are
    quite distinct from those in square, honeycomb and triangular lattices
    studied previously.
   
\end{abstract}
\maketitle
\section{ Introduction }

   The Extended Boson Hubbard model with various  kinds of interactions,
   at various kinds lattices ( bipartite or frustrated ) at various kinds of filling factors
   ( commensurate $ f=p/q $ or in-commensurate ) is described by the following Hamiltonian \cite{boson,sca,gan}:
\begin{eqnarray}
  H  & = &  -t \sum_{ < ij > } ( b^{\dagger}_{i} b_{j} + h.c. ) - \mu \sum_{i} n_{i} + \frac{U}{2}
       \sum_{i} n_{i} ( n_{i} -1 )   \nonumber  \\
      &  +  &  V_{1} \sum_{ <ij> } n_{i} n_{j}  + V_{2} \sum_{ <<ik>> } n_{i} n_{k} + \cdots
\label{boson}
\end{eqnarray}
    where $ n_{i} = b^{\dagger}_{i} b_{i} $ is the boson density and
    $ U, V_{1}, V_{2} $ are onsite, nearest neighbor (nn) and next nearest neighbor (nnn) interactions
    between the bosons. The $ \cdots $ may include further neighbor interactions and
    possible ring-exchange interactions.  For a bipartite lattice, the sign of $ t $ can be changed by changing
    the sign of $ b_{i} $ in one of the two sublattices. But in a frustrated
    lattice, the sign of $ t $ makes a difference.

    It is very important to extend Boson Hubbard model in bipartite
    lattices to frustrated lattices such as triangular, dice and Kagome lattices,
    because of the following motivations:

    (1) For atatoms adsorptions on bare graphite, the preferred adsorption sites form a triangular lattice.
        The phase diagrams of coverage ( the filling factor ) verse temperature
        resulting from the competitions of these energy scales are very diverse and rich \cite{film1,cole}.
        It was believed that Eqn.\ref{boson} may capture the main
        physics of the phenomena.

   (2)  Atomic physicists  are trying to construct an effective
     two dimensional frustrated optical lattices using laser beams and then load either ultra-cold fermion
     or boson atoms at different filling factors on the lattice.
     They may tune the parameters to realize different phases by going through quantum phase transitions
     \cite{ken,cold}.

     (3)  In the hard-core limit $ U \rightarrow \infty $, due to
     the exact mapping between the boson operator and the spin $
     s=1/2 $ operator: $ b^{\dagger}_{i} = S^{+}_{i}, b_{i}= S^{-}_{i}, n_{i}= S^{z}_{i} +1/2 $,
     the boson  model Eqn.\ref{boson} can be mapped to an anisotropic
     $ S= 1/2 $ quantum spin model in an external magnetic field \cite{gan,sca}:
\begin{eqnarray}
    H & = & - 2 t \sum_{<ij>} ( S^{x}_{i} S^{x}_{j} + S^{y}_{i} S^{y}_{j} ) +
     V_{1} \sum_{<ij>}  S^{z}_{i} S^{z}_{j}    \nonumber   \\
       & + &  V_{2} \sum_{ <<ik>> }  S^{z}_{i} S^{z}_{k} - h \sum_{i} S^{z}_{i} + \cdots
\label{spin}
\end{eqnarray}
    where $ h= \mu - 2  V_{1} - 2 V_{2} $ for a square lattice.
    Note that in this Hamiltonian, there is a ferromagnetic coupling in the  $ XY $ spin components and anti-ferromagnetic
    coupling in the $ Z $ spin component. Again, in a bipartite lattice, the
    sign of $ t $ can be changed by changing the sign of $ S^{x}_{i}, S^{y}_{i} $ in one of the two sublattices,
    but keeping $ S^{z}_{i} $ untouched, so Eqn. \ref{spin} is the same as Quantum Heisenberg Antiferromagnet (QHA).
    However, in a frustrated lattice, the sign of $ t $ makes a
    difference, so Eqn. \ref{spin}  is quite different from the QHA.
    The one to one correspondence between physical quantities in boson model and those in spin model
    are the boson density corresponds to the magnetization  $ n \leftrightarrow M $,
    the chemical potential corresponds to the magnetic field $ \mu \leftrightarrow h $, the compressibility
    corresponds to the susceptibility $ \kappa= \frac{ \partial n }{ \partial \mu} \leftrightarrow \chi= \frac{ \partial M }{ \partial h } $.
    The boson number conservation corresponds to the $ U(1) $ rotation around $ \hat{z} $ axis, the superfluid state $ < b_{i} > \neq 0 $
    corresponds to the $ XY $ ordered state $ < S^{+}_{i} > \neq 0 $, the charge ordered state corresponds to the modulation of $ < S^{z} > $.
    The supersolid corresponds to the simultaneous $ < S^{+}_{i} > \neq 0 $ and
    the modulation of $ < S^{z}_{i} > $ \cite{sca,gan}. In the
    hard-core limit, the
    Eqn.\ref{boson} at half filling ( $ q=2 $ ) has the Particle-Hole ( P-H ) symmetry
     $ b_{i}  \leftrightarrow  b^{\dagger}_{i}, n_{i}
    \rightarrow 1-n_{i} $, it can be mapped to Eqn.\ref{spin} in zero magnetic field $ h=0 $ with the Time-reversal
    symmetry $ S^{+}_{i}  \rightarrow  -S^{-}_{i},  S^{z}_{i} \rightarrow - S^{z}_{i} $.
    Eqn.\ref{boson} on triangular lattice at $ q=2 $ is the prototype
    model to study supersolid state with P-H symmetry \cite{gan}.

    The model Eqn.\ref{boson} with only the onsite interaction on square lattice was first studied in Ref.\cite{boson}.
    The effects of long range Coulomb interactions on the transition was studied in \cite{yeboson}.
    Very recently, the most general cases in square lattice
    at generic commensurate filling factors $ f=p/q $ ( $ p, q $ are relative prime numbers ) were
    systematically studied in \cite{pq1}.
    After performing the charge-vortex duality transformation \cite{dual}, the authors in \cite{pq1}
    obtained a dual theory of Eqn.\ref{boson}
    in term of the interacting vortices $ \psi_{a} $ hopping on the dual lattice subject to a fluctuating
    {\em dual} " magnetic field".
    The average strength of the dual " magnetic field "  through a dual plaquette is equal to the boson density $ f=p/q $.
    This is similar to the Hofstadter problem
    of electrons moving in a crystal lattice in the presence of a magnetic field \cite{hof}.
    The magnetic space group (MSG) in the presence of this dual magnetic field
    dictates that there are at least $ q $-fold degenerate minima in the mean field energy spectrum.
    The $ q $ minima can be labeled as $ \psi_{l}, l=0,1,\cdots, q-1 $ which forms a $ q $ dimensional
    representation of the MSG.  In the continuum limit, the final effective theory describing
    the superconductor to the insulator transition in terms of these $ q $ order parameters should be
    invariant under this MSG.  If $  < \psi_{l} > =0 $ for every $ l=\pm $, the system is in the superfluid state.
    If  $ < \psi_{l} > \neq 0 $ for at least one $ l $, the system is in the insulating state.
    In the {\bf supersolid } state \cite{sca,gan,nature}, one condenses a {\bf vortex-antivortex pair},
    but still keeps $  < \psi_{l} > =0 $ for every $ l $.
    In the insulating or supersolid state, there must exist some kinds of charge density wave (CDW)
    ( we assume that every boson carries one internal charge )
    or valence bond solid ( VBS) states which may be stabilized by longer range interactions or possible
    ring exchange interactions included in Eqn.\ref{boson}.
    Very recently, the dual method was used to study the Extended Boson Hubbard model on a triangular lattice \cite{tri}.

   In a recent paper \cite{nature}, one of the authors applied the dual approach of the extended boson Hubbard model
   Eqn.\ref{boson} to study the reentrant "superfluid"  in a narrow region of coverages
   in the second layer of $^{4}He $
   adsorbed on graphite detected by Crowell and Reppy's torsional oscillator experiment in 1993 \cite{he1,he}.
   He showed that there are two consecutive transitions at zero temperature {\em driven by the coverage }: a
    Commensurate-Charge Density Wave (CDW) at half filling to a narrow window of supersolid, then to
    an Incommensurate-CDW. In the Ising limit, the supersolid is a CDW supersolid;
    whereas in the easy-plane limit, it is a  valence bond supersolid.
    Both transitions are second order transition with exact critical exponents $
    z=2, \nu=1/2, \eta=0 $. The results concluded that $^{4}He $ lattice supersolid was
    already observed in 1993.
    He also applied the same dual method to study $ H_2 $/Kr/graphite system investigated in the recent
    experiment \cite{honey} and proposed that a judicious choice of substrate could also lead
    to an occurrence of hydrogen lattice supersolid. Implications to
    the realizations of a lattice supersolid of
    ultra-cold atoms in optical lattices were also given in Ref. \cite{nature}.

    Note that in the dual vortex picture, there are always
    interactions between vortices. Because the phase
    factors from the dual magnetic field only appear in the kinetic
    term, the interactions always
    commute with any generators in the MSG, so will not change the symmetry of the MSG.

    In this paper, we study the Hofstadter bands of vortices hopping
    in the presence of dual magnetic field $ f=p/q $ on the 5 most
    common bipartite and frustrated lattices such as
    square, honeycomb, triangular, dice and Kagome lattices.
    We especially study the bandwidth of the lowest bands.
    There are at least two motivations to study the bandwidth of the lowest
    bands (1) As pointed out in \cite{pq1}, as $ q $ becomes too large, the dual
    vortex method suffer the following two drawbacks (a) As dictated by the MSG,
    there are $ q $ minima in the BZ, so the distance in momentum space between these
    minima scales as $ 1/q $, the continuum theory only works at $ k
    \ll 1/q $, therefore applies only at  distance $ \gg q $. The
    validity regime of the dual vortex theory shrinks.
    (b) When integrating out the vortex modes away from the minima,
    one encounters energy denominators determined by this bandwidth,
    so the dual vortex method may completely break down if the
    bandwidth becomes too small.
    By a simple argument, they estimated that at large $ q $, the bandwidth $ W $ of
    the lowest Hofstadter band scales as $ W \sim e^{-cq}  $ with $ c  $ at
    the order of 1 \cite{cont}. So the
    smaller the bandwidth, the smaller the valid regime of the dual
    vortex approach. (2) In the dual vortex picture, there are
    both a kinetic energy term and interactions between vortices.
    The kinetic term favor the moving of the vortices, while
    the interactions favor the localization of the vortices, the
    competition of the two energy scales may result all kinds of
    phases such as superfluid, CDW, VBS and even supersolid phase \cite{nature}.
    In this paper, we focus on the kinetic term only.
    Calculating the bandwidth of the kinetic term is very important, because the smaller of
    the bandwidth, the more inert the vortices ( the less mobil the vortices are ),
    therefore the boson superfluid state is more likely to occur.

    By choosing suitable gauges and solve corresponding Harper's equations in the 5 lattices.
    we count the number of
    bands, determine the number of minima and their locations in the lowest Hofstadter bands.
    The results are listed in Table 1.
    We also numerically calculate
    the bandwidths of the lowest bands in these
    lattices at any $ q $ and test against the estimate $ W \sim A
    e^{-cq} $. We believe that although the argument in \cite{pq1} seems reasonable, it is far
    from being convincing. So it is important to test this argument
    by quantitative numerical calculations.
    We find that except the Kagome lattice at odd $ q $,
    the exponential law is indeed satisfied  and determine  $ ( A, c ) $ for the 5 different
    lattices. The results are listed in Table 2.
    We find that at given $ q $, the bandwidth $ W $ decreases in the order of
     Triangle, Square and Honeycomb lattice. The corresponding direct lattices
     are honeycomb, square and triangular lattices, so the tendency to form a superfluid
     state increases.
     As shown in the Table 2, when $ q=2 $, the lowest bands in both Dice and Kagome
     lattices are flat. In dice lattice, the gap between the second flat band and the lowest flat band
     is $ \sqrt{6}  $. It indicates that for the original boson at half filling with nearest neighbor hopping on
     the Kagome lattice, there could be only superfluid state.
     However, in Kagome lattice, the gap between the second dispersive  band and the lowest flat band
     vanishes at $ \vec{k}=(0,0) $, so the second dispersive band can not be ignored even in the lowest energy limit.
     Due to the gap vanishing on the Kagome lattice, we can not say definite things about the
     ground state in the original boson on a Dice lattice.
    There are some previous results on the energy spectra on square, honeycome and triangular lattices
    \cite{huse,pq1,nature,tri,un} with different focuses. Our results on dual Dice and Kagome lattices, especially
    the discussions on the possible boson ground states on corresponding direct lattices are new and most
    interesting.

    There are two equivalent methods to be used to study the Hofstadter bands. One is the Magnetic Brillouin
    Zone (MBZ)  method to be employed in the main text.
    This method is physically more transparent and intuitive. Another is
    the symmetric method used in \cite{pq1} and to be used in the appendix.
     This method treat $ x $ and $ y $ coordinates on equal footing, so is
    more symmetric than the first one. In the main text, we will use
    the first method to derive Harper's equations in the 5 lattices
    and then solve the equations analytically at small $ q $ and
    numerically at large $ q $. In the appendix, we will use the
    second method to repeat the calculations.
    Although the coefficients of Harper's equations in the two schemes are different, as expected, we find that
    they result in the same energy spectra.

    In the following, we will first study two bipartite lattices,
    namely square and honeycomb lattices, then we will investigate 3
    frustrated lattices namely triangular, Dice and Kagome lattices.
    In the final section, we summarize our results in Table 1 and
    Table 2, we also comment on the results on CDW formations in
    high temperature superconductors claimed in \cite{pq1} where $ q $ as large as $ 8, 16,
    32 $ are used. In most of the cases, we focus on $ p=1 $ case.

\section{  Square lattice }

  We are looking at the Hofstadter band of vortices hopping around
  square lattice  in the presence of magnetic flux $ f=p/q $ per
  square \cite{pq1} ( Fig.1).
\begin{figure}[ht]
  \centerline{\epsfxsize 0.6\linewidth \epsfbox{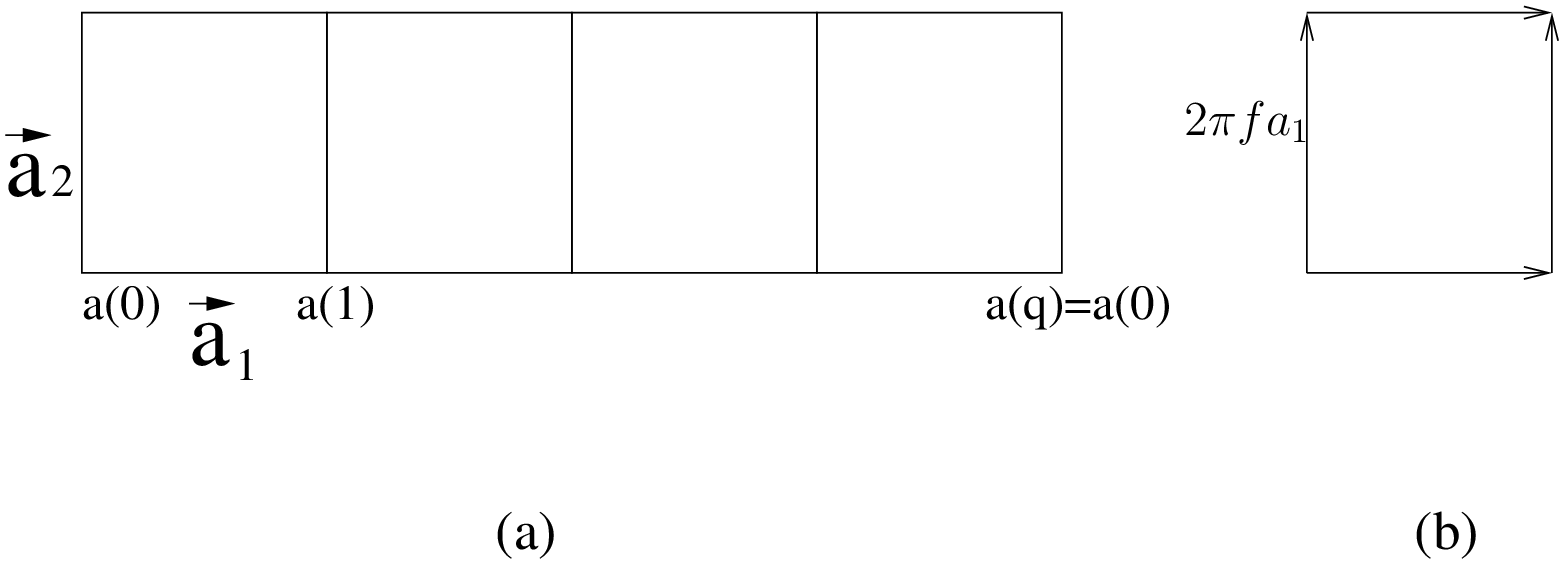}}
  {\footnotesize {\bf Fig 1:} (a) a magnetic unit cell of square
lattice,(b) Phase factors on bonds, 0 phase factors are not shown.}
  \end{figure}
  In the MBZ method, one magnetic unit cell is q times larger than the
  conventional unit cell (Fig.1a). For the simplest gauge chosen in Fig.1(b),
  the hopping Hamiltonian is:
\begin{equation}
H=-t\sum_{\vec{x}}[|\vec{x}+\vec{a_{1}}><\vec{x}|+|\vec{x}+\vec{a_{2}}>e^{i2\pi
fa_{1}}<\vec{x}|+h.c.]
\end{equation}

   In the following, for simplicity, we set $ t=1 $. The eigenvalue equation $ H \psi( \vec{k} )= E(\vec{k} ) \psi( \vec{k}) $ leads to the Harper's equation:
\begin{eqnarray}
&&-e^{-ik_{x}}\psi_{l-1}(\vec{k})-2\cos(2\pi fl+k_{y})\psi_{l}(\vec{k})-e^{ik_{x}}\psi_{l+1}(\vec{k})=E(\vec{k})\psi_{l}(\vec{k})
\label{square}
\end{eqnarray}
    where $ l=0,\cdot \cdot \cdot, q-1; {-\pi\over q}\leq k_{x} \leq {\pi\over
    q}; -\pi \leq k_{y} \leq \pi $.

  For small values of $ q $, Eqn.\ref{square} can be solved
  analytically. For large values of $ q  $, we solve it numerically. There are always $ q $ bands.
  We focus on the lowest energy band and its bandwidth.
  As shown in \cite{pq1}, there are q minima at ${(0,2\pi f l),l=0,\cdot\cdot\cdot,q-1}$.
  The spectrum for $ q=4 $ is shown in Fig. 2. In order to see
  clearly all the 4 MSG related minima, only part of the energy band close to
  the 4 minima in the lowest band is included.
\begin{figure}[ht]
  \centerline{\epsfxsize 0.8\linewidth \epsfbox{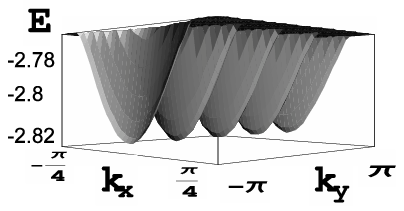}}
{\footnotesize {\bf Fig 2:} The lowest energy band of square lattice
   at $ q=4 $. }
\vspace{0.15cm}
\end{figure}
   We also numerically calculated the bandwidth of the lowest band
   upto $ q=18 $. We found that it indeed satisfy the exponential law $
   W=A e^{-cq} $ with $ A=26.05, c=-1.20 $. In a semi-log plot, it is
   a straight line which is shown in Fig.3

\vspace{0.25cm}

\begin{figure}[ht]
  \centerline{\epsfxsize 0.4\linewidth \epsfbox{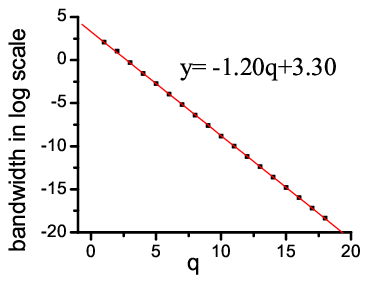}}
\vspace{0.15cm}
{\footnotesize {\bf Fig 3:} The bandwidth of lowest band in square lattice vs q}
\end{figure} 
\vspace{0.25cm}

    What is surprising is that even for the smallest $ q=1 $ which
    is  the no magnetic field case, the exponential law is still
    satisfied.

\section{ Honeycomb lattice}

     Honeycomb lattice is not a Bravais lattice, it can be thought as a underlying  parallelogram Bravais lattice with
     two primitive vectors $ \vec{a}_{1}=  \hat{x}, \vec{a}_{2}= \frac{1}{2} \hat{x} + \frac{ \sqrt{3} }{ 2 } \hat{y} $
     plus a two point basis located at $ \vec{x} +  \vec{\delta} $ and
     $ \vec{x} + 2 \vec{\delta} $ where $ \vec{ \delta}= \frac{1}{3}( \vec{a}_{1} + \vec{a}_{2} ) $ ( Fig.2 ).
     Its reciprocal lattice is also a parallelogram Bravais lattice spanned by $ \vec{k}= k_{1} \vec{b}_{1}
     + k_{2} \vec{b}_{2} $ with $ \vec{b}_{i} \cdot \vec{a}_{j} = \delta_{ij} $.

\vspace{0.25cm}
\epsfig{file=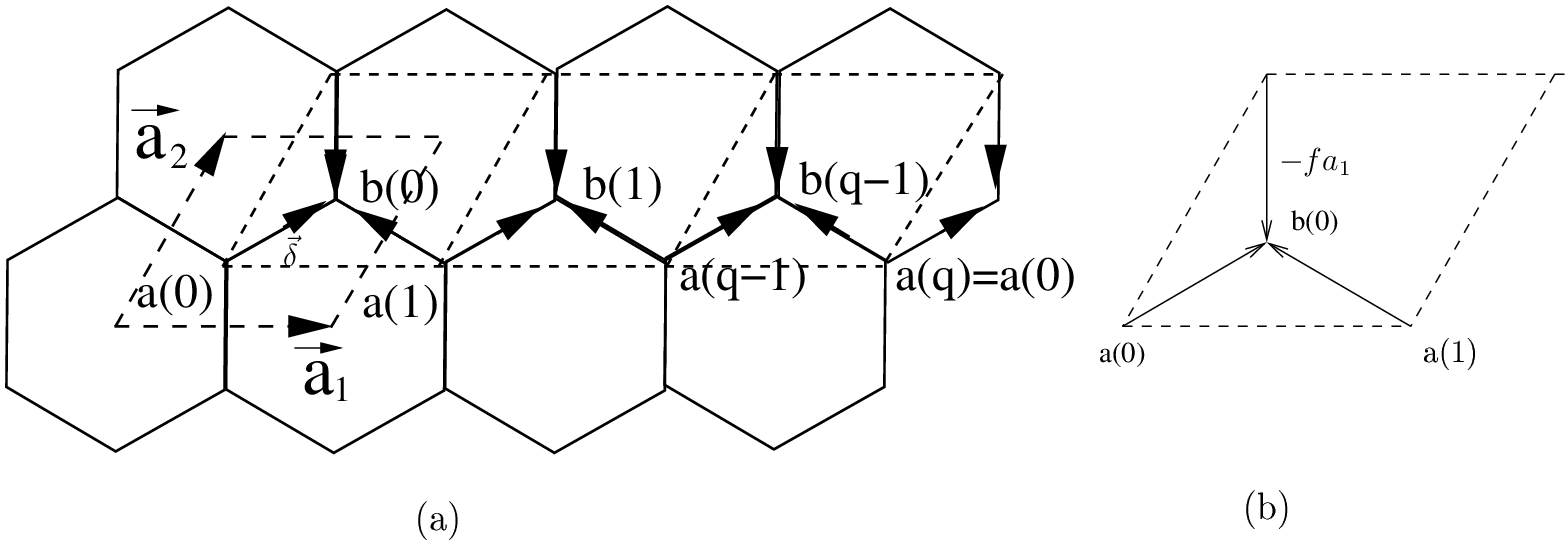,width=6in,height=1.8in,angle=0}
\vspace{0.25cm}

{\footnotesize {\bf Fig 4:} (a) A magnetic unit cell of honeycomb
lattice (b) Phase factors on bonds, 0 phase factors are not shown }

\vspace{0.25cm}

  In the MBZ method, one magnetic unit cell is q times larger than the
  conventional unit cell (Fig.4a). In one
  conventional unit cell, there are also two atoms which are labeled by two color indices $ a$ and $b $
  ( Fig.4).
  We are looking at the Hofstadter band of vortices hopping around
  a honeycomb lattice in the presence of magnetic flux $ f=p/q $ per
  hexagon. For the simplest gauge chosen in Fig. 4(b), the vortex hopping Hamiltonian is:
\begin{eqnarray}
H&=&-t\sum_{\vec{x}}[|\vec{x}+\vec{\delta}><\vec{x}|+|\vec{x}+\vec{\delta}><x+\vec{a_{1}}|\nonumber \\
&&+|\vec{x}+\vec{\delta}>e^{-i2\pi fa_{1}}<\vec{x}+\vec{a_{2}}|+h.c.]
\end{eqnarray}

  The Harper's equation is:
\begin{eqnarray}
&&-(1+e^{i(k_{x}+2\pi fl)})\psi^{a}_{l}(\vec{k})-e^{ik_{y}}\psi^{a}_{l+1}(\vec{k})=E(\vec{k})\psi^{b}_{l}(\vec{k})\nonumber\\
&&-(1+e^{-i(k_{x}+2\pi fl)})\psi^{b}_{l}(\vec{k})-e^{-ik_{y}}\psi^{b}_{l-1}(\vec{k})=E(\vec{k})\psi^{a}_{l}(\vec{k})
\label{honey}
\end{eqnarray}
    where $ l=0,\cdot \cdot \cdot, q-1 $ is the flavor indices and $
    a,b $ is the color indices, $ -{\pi\over q}\leq k_{x} \leq {\pi\over q} $ .

  For small values of $ q $, Eqn.\ref{honey} can be solved
  analytically.  When ${q=1}$, there is actually no magnetic field,
  it is just ordinary tight-binding model. There are two bands:
  ${\pm \sqrt{3+2(\cos k_{x}+\cos k_{y}+\cos(k_{x}+k_{y}))}}$. The lowest energy band is
  ${-\sqrt{3+2(\cos k_{x}+\cos k_{y}+\cos(k_{x}+k_{y}))}}$. There is only one minimum at ${(0,0)}$.
  The bandwidth is ${3}$. The $ q=2 $ case is especially interesting, because the original boson model can be mapped to a
   quantum $ s=1/2 $ spin model  Eqn.\ref{spin} in a triangular lattice at zero field.
   For $ q=2 $, there are $ 4 $  bands $ E( \vec{k} ) = \pm  t \sqrt{ 3 \pm  \sqrt{ 2 A( \vec{k} ) } } $
   where $ A( \vec{k} ) = 3 + (  \cos 2k_{1} + \cos 2 k_{2} - \cos ( 2 k_{1} - 2 k_{2} ) ) $.
   The lowest subband is $ E( \vec{k} ) = -  t \sqrt{ 3 + \sqrt{ 2 A( \vec{k} ) } } $.
   There are $ 4 $  minima at $ \pm ( \pi/6, -\pi/6 ) $ and $ \pm ( \pi/6, 5 \pi/6 ) $.
   The  $ 4 $ minima transforms to each other under the MSG.

    For general $ q $, there are always $ 2q $ bands. As shown in \cite{tri,un}, there are two cases
    (1) ${q}$ is odd, there are ${q}$ minima at ${(0,2\pi fl)},l=0,\cdot\cdot\cdot,q-1$.
    (2) ${q}$ is even, there are ${2q}$ minima at ${({ \alpha \pi\over3q},-{\alpha \pi\over3q}+2\pi f l)}$
     where $ \alpha=\pm, l=0,\cdot\cdot\cdot,q-1$.

   For large values of $ q $, we solve Eqn.\ref{honey} numerically.
   Just like in square lattice, we focus on the energy band near the minima in the lowest energy band.
   The $ q=3 $ and $ q=4 $  spectrum are shown in Fig. 5(a), 5(b) respectively.
\vspace{0.25cm}
\epsfig{file=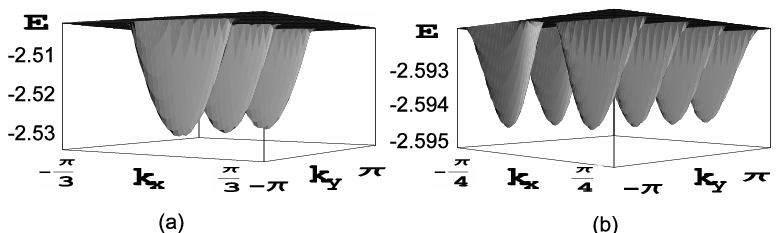,width=6.5in,height=2in,angle=0}
\vspace{0.25cm} {\footnotesize {\bf Fig 5:} The lowest energy bands
of honeycomb lattice at (a) ${q=3}$, (b)${q=4}$ }

   We also numerically calculated the bandwidths of the lowest band
   upto $ q=18 $. We found that they satisfy the exponential law $
   W=A e^{-cq} $ with $ A=11.82, c=-1.66 $ for both $ q $ even and odd. In a semi-log plot, it is
   a straight line which is shown in Fig.6.
\begin{figure}[ht]
  \centerline{\epsfxsize 0.4\linewidth \epsfbox{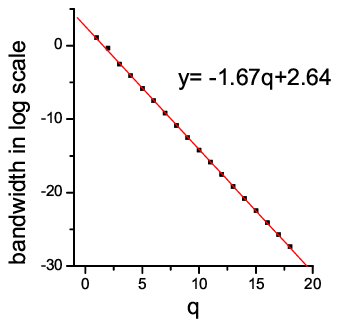}}
 {\footnotesize {\bf Fig 6:} The bandwidth of
    honeycomb lattice vs q}
\end{figure}
\vspace{0.25cm}

    What is surprising is that even for the smallest $ q=1 $ which
    is the no magnetic field case, the exponential law is still
    satisfied.

\section{  Triangular lattice }

   In the previous two sections, we studied two bipartite
   lattices. In this section, we study the simplest frustrated
   lattice which is the triangular lattice. As said in the
   introduction, the physics in frustrated lattices could be very
   different from that in bipartite lattices.
\vspace{0.25cm}
\epsfig{file=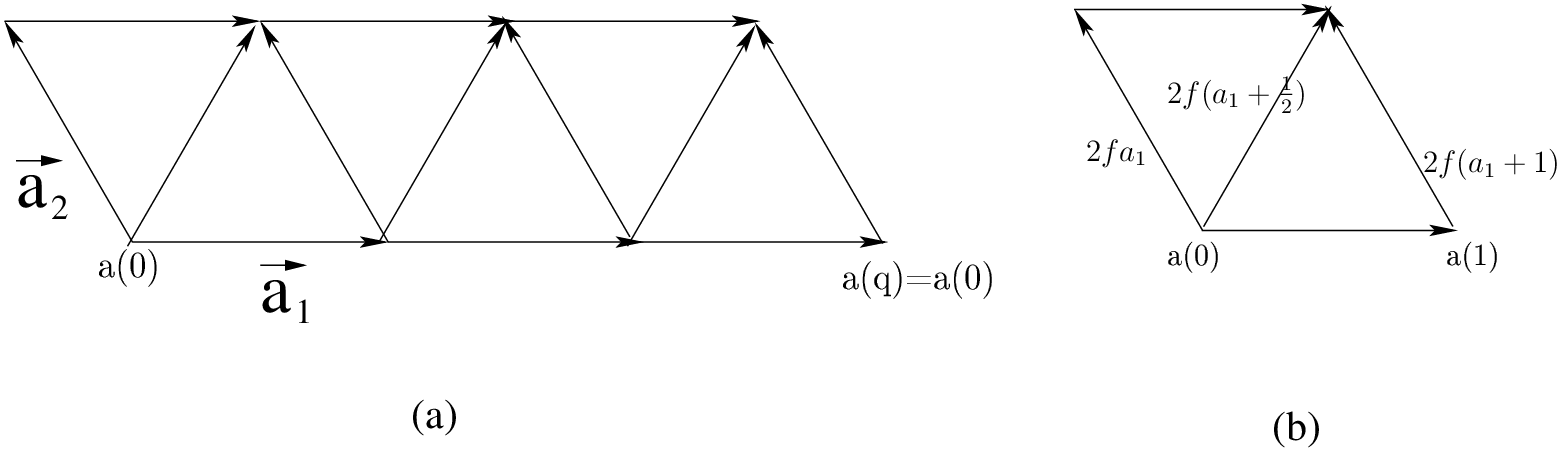,width=6.5in,height=1.8in,angle=0}
\vspace{0.25cm} {\footnotesize {\bf Fig 7:}Triangular Lattice (a)
magnetic unit cell of Triangular lattice, (b)  Phase factors on
bonds, 0 phase factors are not shown.}
\vspace{0.25cm}

    We are looking at the Hofstadter band of vortices hopping around
    a triangular lattice in the presence of magnetic flux $ f=p/q $ per
    triangle. For the simplest gauge chosen in Fig. 7b, the Hamiltonian is:
\begin{eqnarray}
H&=&-t\sum_{\vec{x}}[|\vec{x}+\vec{a_{1}}><\vec{x}|+|\vec{x}+\vec{a_{2}}>e^{i2\pi2 fa_{1}}<\vec{x}|\nonumber\\
&&+|\vec{x}+\vec{a_{1}}+\vec{a_{2}}>e^{i2\pi2f(a_{1}+{1\over{2}})}<\vec{x}|+h.c.]
\end{eqnarray}

     The corresponding Harper's equation is:
\begin{eqnarray}
&&-2\cos(k_{y}+2\pi fl)\psi_{l}(\vec{k})-(e^{-ik_{x}}+e^{-i(k_{x}+k_{y}+2\pi f(2l-1))})\psi_{l-1}(\vec{k})\nonumber\\&&-(e^{ik_{x}}+e^{i(k_{x}+k_{y}+2\pi f(2l+1))})\psi_{l+1}(\vec{k})
=E(\vec{k})\psi_{l}(\vec{k})
 \label{tri}
\end{eqnarray}
   where $ l=0,\cdot \cdot \cdot, q-1 $.

    From the phase factors on the bond, it is easy to see that when ${q}$ is
    even, there are only ${q\over2}$ unit cells in one magnetic unit cell.
    Since the magnetic unit cell shrink to ${q\over 2}$, the range of ${k_{x}}$ in the momentum space
    double its range accordingly. Therefore in Eqn.\ref{tri}, for q is odd,
    ${-{\pi\over q}\leq k_{x} \leq {\pi\over q}}$, while for $ q $
    even, ${-{2\pi\over q}\leq k_{x} \leq {2\pi\over q}}$.

    In fact, as shown in \cite{pq1}, there are three cases in a triangular lattice:
     (1) When ${q}$ is odd, there are $ q $ bands. There are ${q}$ minima at
     ${(0,4\pi fl)},l=0,\cdot\cdot\cdot,q-1$ in the lowest band.
     (2) When ${q}$ is even, there are $ q/2 $ bands. There are still two subcases:
    (2a). $ q=2n$ with $ n $ odd,
    there are ${q}$ minima at ${({2\pi\alpha\over 3q},{2\pi\alpha\over 3q}+4\pi fl)}$,
    ${\alpha=\pm}$ and ${l=0,\cdot\cdot\cdot {q\over 2}-1}$.
    (2b). $ q=2n $ with $ n $ even, there are ${q\over2}$ minima at ${(0,4\pi fl)},l=0,\cdot\cdot\cdot,{q\over2}-1 $.

   For ${q=1}$ which is the no magnetic case, the energy spectrum is
   $ E( \vec{k} )= {-2(\cos k_{x}+\cos k_{y}+\cos(k_{x}+k_{y}))}$, there is only one minima at
   ${(0,0)}$. For ${q=2}$, as shown in \cite{nature}, the spectrum is
   $ E(\vec{k})= {-2(\cos k_{x}+\cos k_{y}-\cos(k_{x}+k_{y}))}$. There are two minima located at
   ${(\pm{\pi\over3},\pm{\pi\over3})}$ \\

   For large ${q}$, we solve Eqn.\ref{tri} numerically. The results
   for $ q=3 $ ( odd case ), $ q=6 $ ( $ 2n $ with $ n $ odd case ) and $ q=8 $ ( $ 2n $ with $ n $ even case
   ) are shown in Fig. 8a,b,c respectively.

\vspace{0.25cm}
\epsfig{file=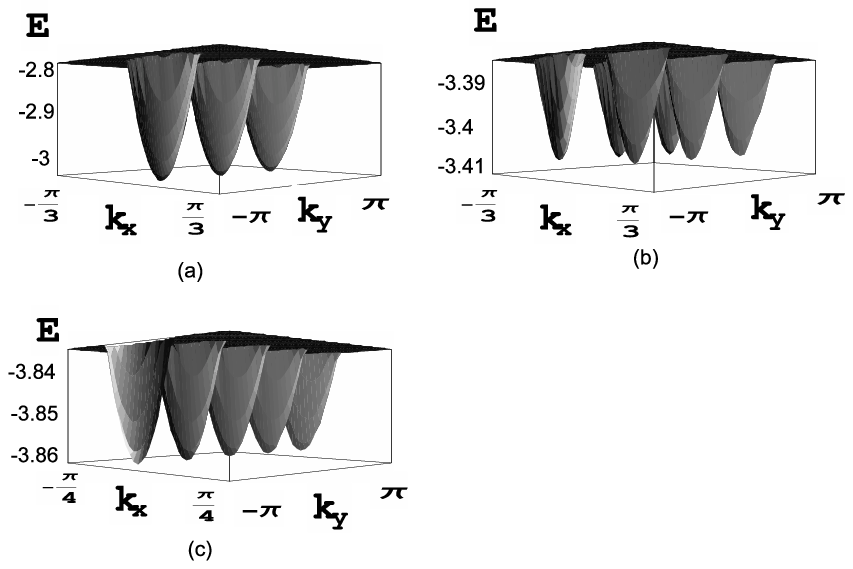,width=6in,height=3in,angle=0}
\vspace{0.25cm} {\footnotesize {\bf Fig 8:} The lowest energy bands
  of triangular lattice at (a) ${q=3}$, (b)${q=6}$, (c)${q=8}$}
\vspace{0.25cm}

   We also numerically calculated the bandwidth of the lowest band
   upto $ q=25 $ for $ q $ is odd and upto $ q=30 $ for $ q $ is even.
   For $ q $ odd, we find $ A=9.21, c=0.82  $ ( Fig. 9a). For $ q=2n $,  both $ n $ is odd and even, the bandwidth
   satisfy the same exponential law with $ A=55.70, c=0.83 $ ( Fig.9b ).

\vspace{0.25cm}
\epsfig{file=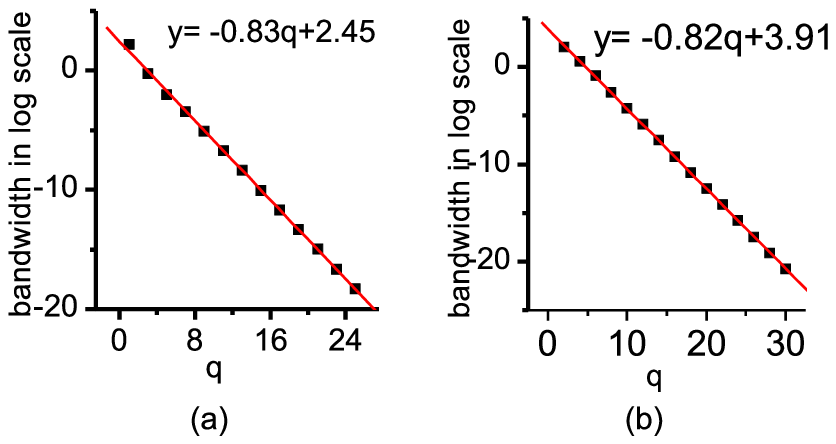,width=6in,height=2in,angle=0}
\vspace{0.25cm} {\footnotesize {\bf Fig 9:} The bandwidth of
 triangular lattice vs q (a) ${q}$ is odd,(b) ${q}$ is even}
\vspace{0.25cm}

\section{ Dice lattice  }

   The dice lattice is the dual lattice of the Kagome lattice.
   It can be thought of consisting of two interpenetrating honeycomb lattice.
   Obviously, the dice lattice is not a Bravais lattice,
   it can be thought as a underlying  parallelogram Bravais lattice with
   two primitive lattice vectors $ \vec{a}_{1}=  \hat{x}, \vec{a}_{2}= \frac{1}{2} \hat{x} +
   \frac{ \sqrt{3} }{ 2 } \hat{y} $ plus a three point basis labeled $ a, b, c $ located at $ \vec{x},
   \vec{x} +  \vec{\delta}, \vec{x} + 2 \vec{\delta} $ where $ \vec{ \delta}= \frac{1}{3}(
   \vec{a}_{1} + \vec{a}_{2} ) $ ( Fig.10).
   In contrast to the honeycomb lattice shown in Fig.4, the dice lattice is not a bipartite lattice and
   has a 3-sublattice structure.

\vspace{0.25cm}
\epsfig{file=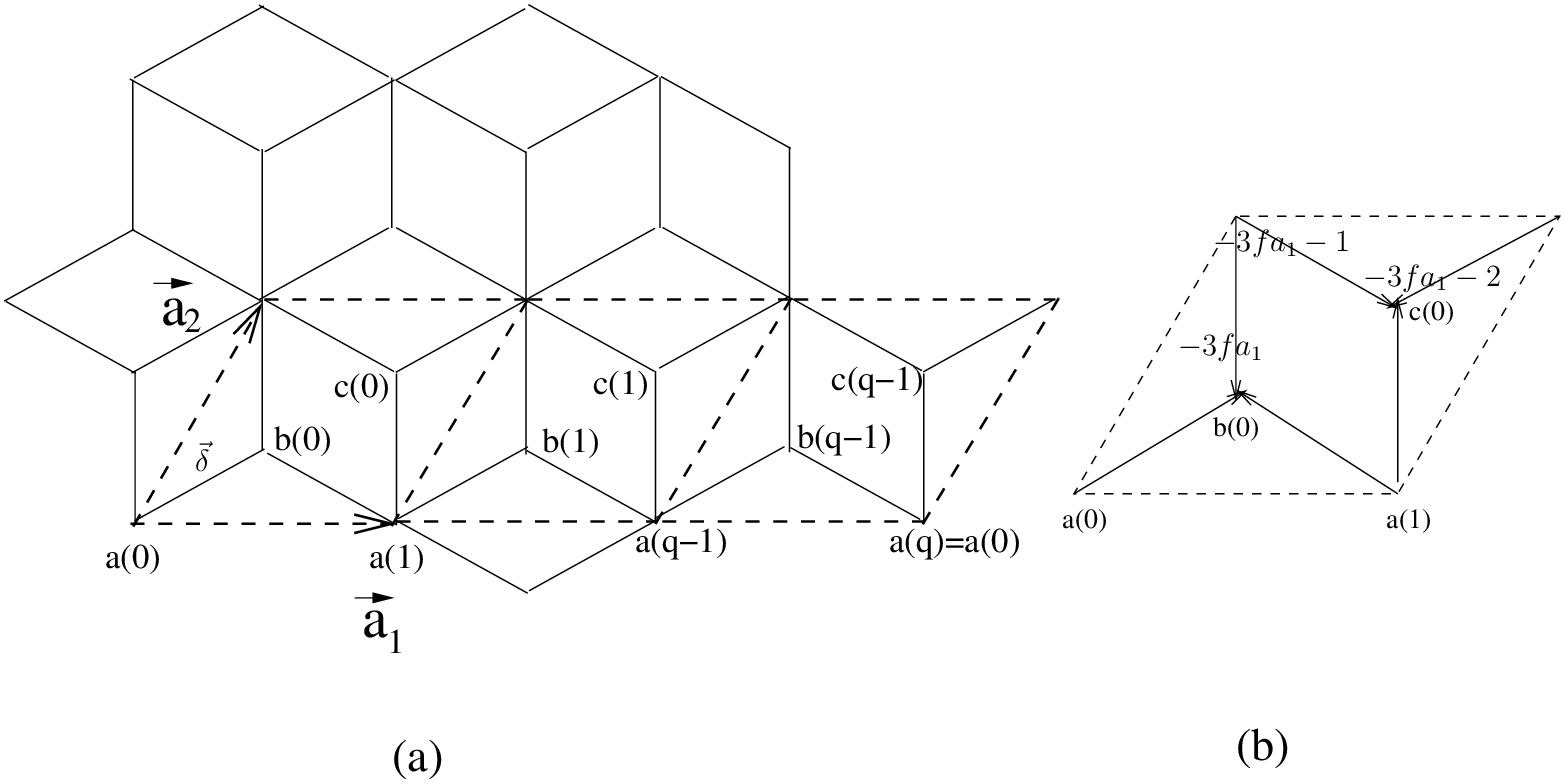,width=6in,height=2.4in,angle=0}
\vspace{0.25cm} {\footnotesize {\bf Fig 10:}Dice Lattice (a)
magnetic unit cell of Dice lattice, (b) Phase factors on bonds, 0
phase factors are not shown.}

\vspace{0.25cm}

    We are looking at the Hofstadter band of vortices hopping around
    a dice lattice in the presence of magnetic flux $ f=p/q $ per
    parallelogram.  For the simplest gauge chosen in Fig 10b, the Hamiltonian is:
\begin{eqnarray}
H&=&-t\sum_{\vec{x}}[|\vec{x}+\vec{\delta}><\vec{x}|+|\vec{x}+\vec{\delta}>e^{-i2\pi3fa_{1}}<\vec{x}+\vec{a_{2}}|+|\vec{x}+\vec{\delta}><\vec{x}+\vec{a_{1}}|\nonumber\\
&&+|\vec{x}+\vec{2\delta}>e^{-i2\pi3f(a_{1}+{1\over3})}<\vec{x}+\vec{a_{2}}|+|\vec{x}+\vec{2\delta}><\vec{x}+\vec{a_{1}}|\nonumber\\  &&+|\vec{x}+\vec{2\delta}>e^{-i2\pi3f(a_{1}+{2\over3})}<\vec{x}+\vec{a_{1}}+\vec{a_{2}}|+h.c.]
\end{eqnarray}

    The corresponding Harper's equation is
\begin{eqnarray}
&&-(1+e^{-i(k_{y}+2\pi3fl)})\psi^{b}_{l}(\vec{k})-e^{-ik_{x}}\psi^{b}_{l-1}(\vec{k})-e^{-i(k_{y}+2\pi3f(l+{1\over3}))}\psi^{c}_{l}(\vec{k})\nonumber \\
&&-(e^{-ik_{x}}+e^{-i(k_{x}+k_{y}+2\pi3f(l+{2\over3}))})\psi^{c}_{l-1}(\vec{k})=E(\vec{k})\psi^{a}_{l}(\vec{k});\nonumber\\
&&-(1+e^{-i(k_{y}+2\pi3fl)})\psi^{a}_{l}(\vec{k})-e^{ik_{x}}\psi^{a}_{l+1}(\vec{k})=E(\vec{k})\psi^{b}_{l}(\vec{k});\nonumber\\
&&-e^{i(k_{y}+2\pi3f(l+{1\over3}))}\psi^{a}_{l}(\vec{k})-(e^{i(k_{x}+k_{y}+2\pi3f(l+{2\over3}))}+e^{ik_{x}})\psi^{a}_{l+1}(\vec{k})\nonumber \\
&&=E(\vec{k})\psi^{c}_{l}(\vec{k})
\end{eqnarray}
   where $ l=0,1,\cdots,q-1 $ is the flavor indices and $ a,b,c $ are
   the 3 color indices.

     For the simplest gauge shown in Fig.10b, we need to distinguish two general cases: ${q=3n}$ and ${q\neq3n}$.
     For ${q\neq 3n}$, we find out there are still two subcases: q is even and  q  is  odd.

    When q is small, we can solve the Harper's equation analytically.
    For ${q=1}$ which is the no magnetic field case, there are 3 bands:
    ${\pm \sqrt{3+2(\cos k_{x}+\cos k_{y}+\cos(k_{x}-k_{y}))}}$ and 0. The lowest band is
    $ E( \vec{k}) ={-\sqrt{3+2(\cos k_{x}+\cos k_{y}+\cos(k_{x}-k_{y}))}}$.
    The minimum is at ${(0,0)}$.
    For ${q=2}$, all the 6 bands are completely flat. The energies are
    $ E=\sqrt{6},0, -\sqrt{6} $, each with degeneracy $ 2 $.
    For ${q=3}$, there are also 3 bands: ${\pm\sqrt{6+2A(k_{x},k_{y})}}$ and 0. The lowest band is $ E( \vec{k}) =
    {-\sqrt{6+2A(k_{x},k_{y})}}$ where $
    A(k_{x},k_{y})=\cos k_{x}+\cos k_{y}+\cos(k_{x}-k_{y})+\cos(k_{y}+{4\pi\over 3}) +
    \cos(k_{y}-k_{x}+{2\pi\over 3})+\cos(k_{x}+{2\pi\over 3})  $. The  two minima are at
     ${(0,0)}$ and ${(-{2\pi\over 3},{2\pi\over 3})}$.

\vspace{0.25cm}
\epsfig{file=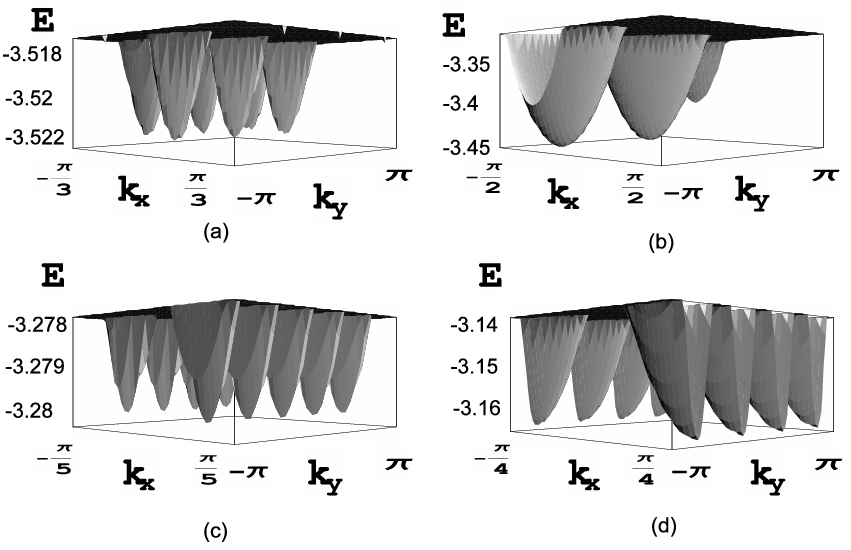,width=6in,height=2.5in,angle=0}
\vspace{0.25cm} {\footnotesize {\bf Fig 11:} The lowest energy bands
 of dice lattice at (a) ${q=9}$, (b)${q=6}$, (c)${q=5}$, (d)${q=4}$}

\vspace{0.25cm}

\vspace{0.25cm}
\epsfig{file=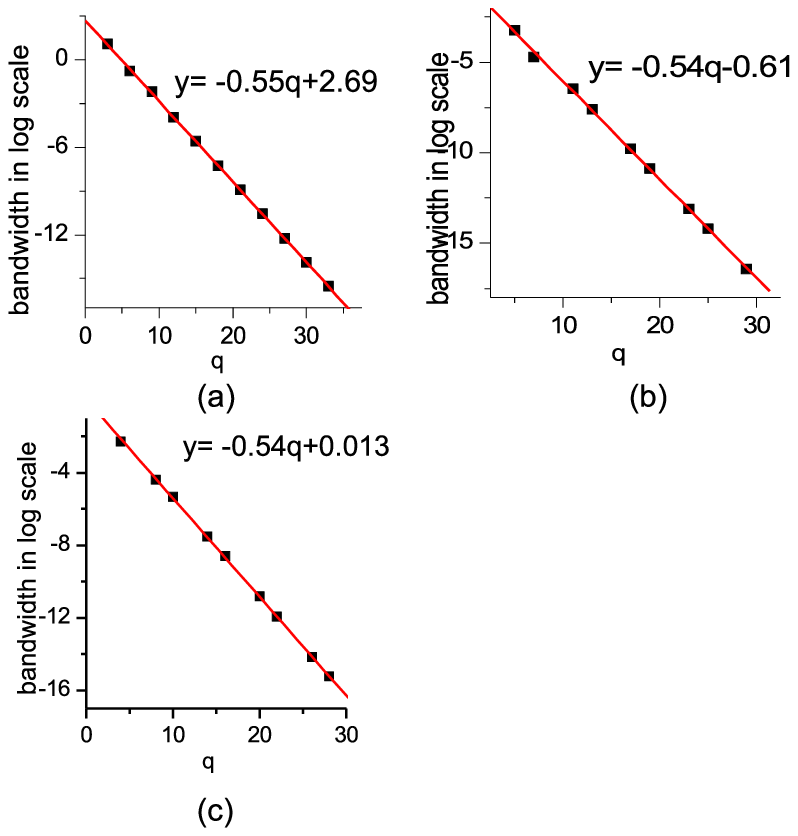,width=6in,height=3in,angle=0}
\vspace{0.25cm} {\footnotesize {\bf Fig 12:} The bandwidths of dice
lattice vs q (a) ${q=3n}$,(b) ${q\neq 3n}$ and odd,(c) ${q\neq 3n}$
and even }
\vspace{0.25cm}

  In general,there are four cases in the dice lattcie:
  (1)  ${q=3n}$,${{-3\pi\over q}\leq k_{x} \leq {3\pi\over q}} $.
  There are $ q $ bands.
  We also need to distinguish two subcases: (1a). ${n}$ is odd, there are ${2n}$ minina at
  ${(0,{2\pi\over n}l)}$ and ${(-{2\pi\over 3n},{2\pi\over
  3n}+{2\pi\over n}l)},l=0,\cdot\cdot\cdot,n-1$. $ q=9$ case is shown in Fig.11(a). (1b). ${n}$
  is even, there are ${n}$ minima at ${(-{\pi\over3n},{2\pi\over
  3n}+{2\pi\over n}l)}, l=0,\cdot\cdot\cdot,n-1$. $ q=6 $ case is shown in Fig.11(b). For
  both cases, the bandwidth falls as ${14.73e^{-0.55q}}$ as shown in Fig.12(a).
  (2)  ${q \neq 3n}$,${{-\pi\over q}\leq k_{x} \leq {\pi\over q}}$.
  There $ 3q $ bands. We also need to distinguish two subcases:
   (2a). ${q}$ is odd, there are ${2q}$ minima at ${({2\alpha\pi\over 3q},-\pi+{\alpha\pi\over 3q}+{2\pi\over q}l)}$
   $\alpha=\pm,l=0,\cdot\cdot\cdot,q-1$. $ q=5 $ case is shown in
   Fig.11(c). The bandwidth falls as ${0.54e^{-0.54q}}$ as shown in
   Fig.12(b).
   (2b). ${q}$ is even, there are ${q}$ minima at ${({\pi\over q},{\pi\over q}+{2\pi\over q}l)},
   l=0,\cdot\cdot\cdot,q-1$. $ q=4 $ case is shown in Fig.11(d), the bandwidth falls as ${1.01e^{-0.54q}}$ as shown
   in Fig.12c.

   It seems to us that all the four cases have the same $ c $ within numerical errors,
   but with different magnitudes $ A $.

\section{ Kagome lattice }

  Kagome lattice is not a Bravais lattice either,
  it can be thought as a underlying  parallelogram Bravais lattice with
  two primitive lattice vectors $ \vec{a}_{1}=  \hat{x}, \vec{a}_{2}= \frac{1}{2} \hat{x} + \frac{ \sqrt{3} }{ 2 } \hat{y} $
  plus a three point basis labeled $ a, b, c $ located at $ \vec{x},  \vec{x} +  \vec{a}_{1}/2,
  \vec{x} +  \vec{a}_{2}/2 $ as shown  in Fig.13. Note that the
  Kagome lattice contains both triangles and hexagons.

\vspace{0.25cm}
\epsfig{file=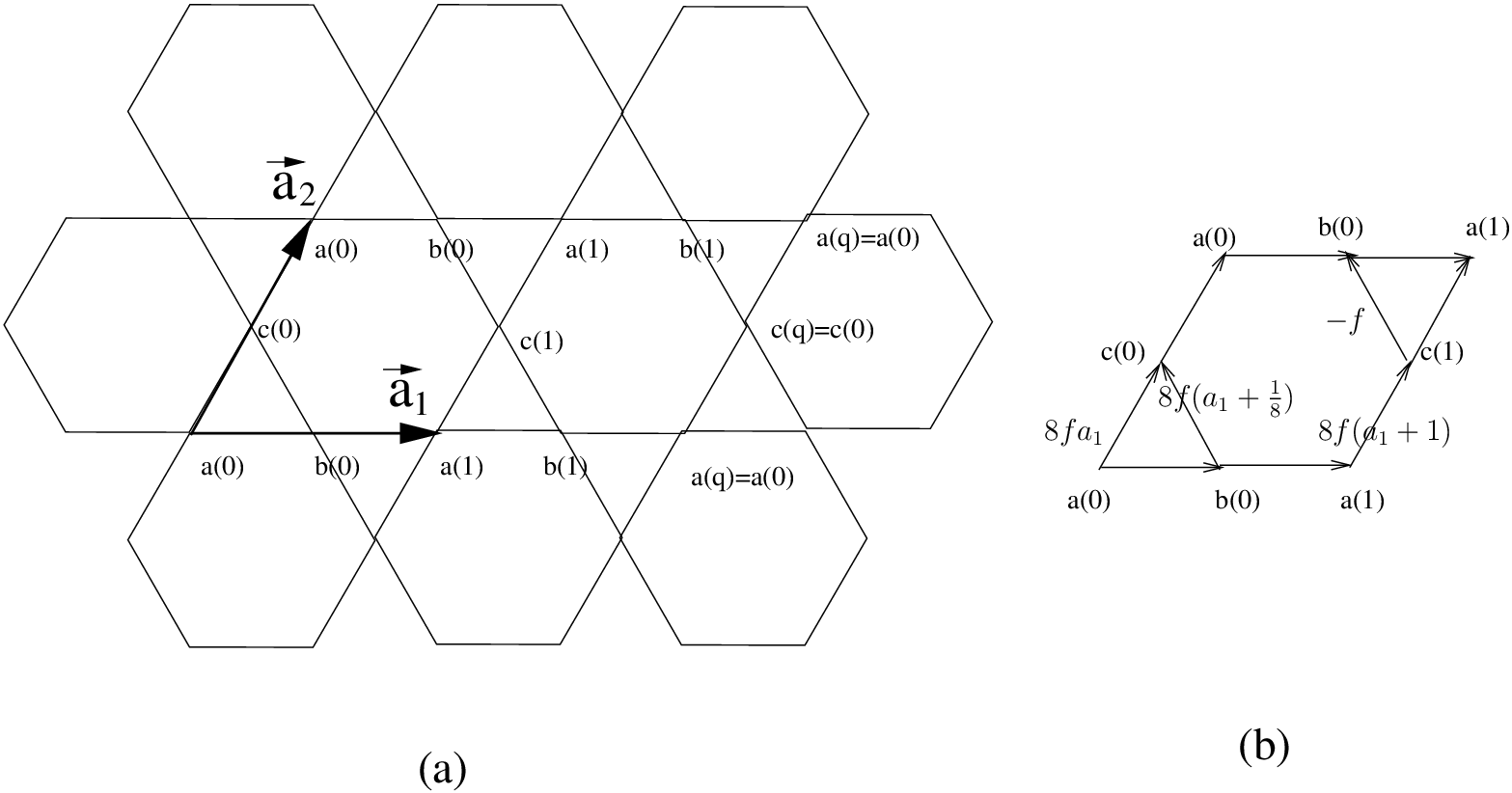,width=6in,height=2.6in,angle=0}
\vspace{0.25cm} {\footnotesize {\bf Fig 13:} Kagome
 lattice (a) magnetic unit cell of Kagome lattice, (b) Phase factors on bonds, 0 phase factors are not shown. }
\vspace{0.25cm}

    We are looking at the Hofstadter band of vortices hopping around
    a Kagome lattice in the presence of magnetic flux $ f=p/q $ per
    triangle and $6f $ flux quantum per hexagon. So overall, there are $ 8f $ flux quanta per
    parallelogram. For the simplest gauge chosen in Fig 13b, the Hamiltonian is:

\begin{eqnarray}
H&=&-t\sum_{\vec{x}}[|\vec{x}+\vec{{a_{1}\over2}}><\vec{x}|+|\vec{x}+\vec{{a_{2}\over2}}>e^{i2\pi8fa_{1}}<\vec{x}|+|\vec{x}+\vec{a_{1}}><\vec{x}+\vec{{a_{1}\over2}}|\nonumber\\
&&+|\vec{x}+\vec{a_{2}}><\vec{x}+\vec{{a_{2}\over2}}|+|\vec{x}+\vec{{a_{1}\over2}}+\vec{a_{2}}>e^{-i2\pi f}<\vec{x}+\vec{a_{1}}+\vec{{a_{2}\over2}}|\nonumber\\
&&+|\vec{x}+\vec{{a_{2}\over2}}>e^{i2\pi8f(a_{1}+{1\over8})}<\vec{x}+\vec{{a_{1}\over2}}|+h.c.]
\end{eqnarray}

   The corresponding Harper's equation is:
\begin{eqnarray}
&&-\psi^{b}_{l}(\vec{k})-e^{ik_{x}}\psi^{b}_{l-1}(\vec{k})-(e^{ik_{y}}+e^{-i2\pi8fl})\psi^{c}_{l}(\vec{k})=E(\vec{k})\psi^{a}_{l}(\vec{k});\nonumber\\
&&-\psi^{a}_{l}(\vec{k})-e^{-ik_{x}}\psi^{a}_{l+1}(\vec{k})-e^{-i2\pi8f(l+{1\over8})}\psi^{c}_{l}(\vec{k})-e^{-i(2\pi f+k_{x}-k_{y})}\psi^{c}_{l+1}(\vec{k})\nonumber\\
&&=E(\vec{k})\psi^{b}_{l}(\vec{k});\nonumber\\
&&-(e^{-ik_{y}}+e^{i2\pi8fl})\psi^{a}_{l}(\vec{k})-e^{i2\pi8f(l+{1\over8})}\psi^{b}_{l}(\vec{k})\nonumber\\
&&-e^{i(2\pi
f+k_{x}-k_{y})}\psi^{b}_{l-1}(\vec{k})=E(\vec{k})\psi^{c}_{l}(\vec{k})
\end{eqnarray}

\vspace{0.25cm}
\epsfig{file=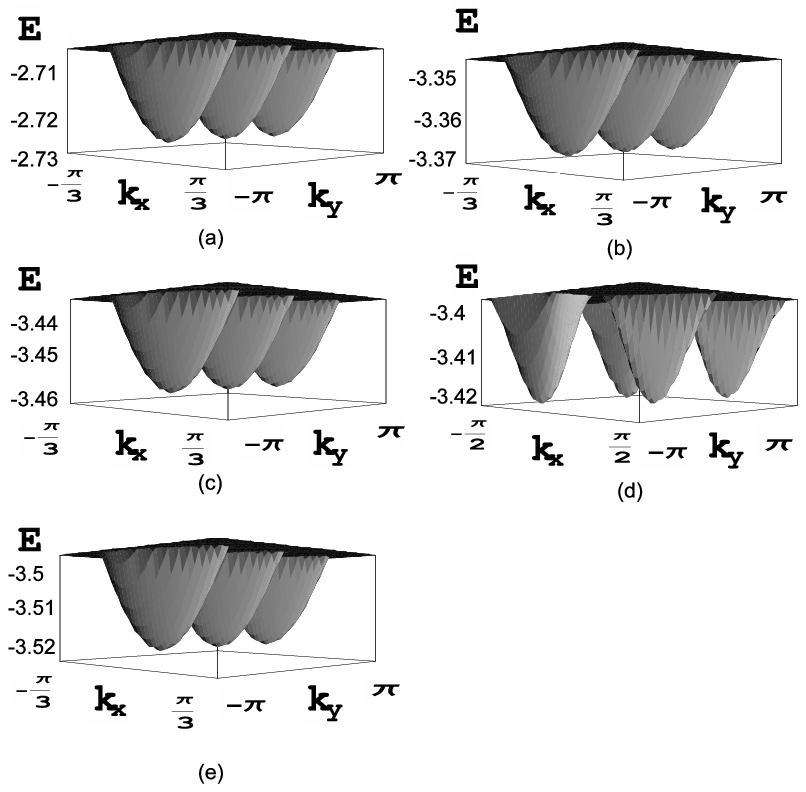,width=6in,height=3.0in,angle=0}
\vspace{0.25cm} {\footnotesize {\bf Fig 14:} The lowest energy bands
   of Kagome lattice at (a) ${q=3}$, (b) ${q=6}$, (c) ${q=12}$,(d) ${q=16}$,(e)
   ${q=24}$}.
\vspace{0.25cm}\\

 For the gauge chosen in Fig.13a, we can solve the spectra at ${q=1,2,4,8}$ exactly, because
 for all the four cases, we only need to solve a 3 by 3 matrix whose
 secular equation is a cubic equation $ {\lambda^3-(4+A(k_{x},k_{y}))\lambda+2\cos({2\pi\over q})
 A(k_{x},k_{y})=0}$ where $ {A(k_{x},k_{y})=2+2(\cos k_{x}+\cos
 k_{y}+\cos(k_{x}-k_{y}))} $. There are 3 bands.  For ${q=1}$ which is the non-magnetic case,
 the 3 bands  are${-1\pm \sqrt{1+A(k_{x},k_{y})}, 0 }$.
 The lowest band is $ E( \vec{k})= {-1-\sqrt{1+A(k_{x},k_{y})}}$  whose
 minimum is at ${(0,0)}$.
 For ${q=2}$, the 3 bands are ${-2, 1-\sqrt{1+A(k_{x},k_{y})},
 1+\sqrt{1+A(k_{x},k_{y})}}$.
 We can see the lowest band is completely flat, the second band
 touches the lowest band at $ \vec{ k } =(0,0) $ where the gap vanishes !
 For ${q=4}$, the 3 bands are $ \pm \sqrt{4+A(k_{x},k_{y})}, 0 $. The lowest band is
 $ E(\vec{k} )= {-\sqrt{4+A(k_{x},k_{y})}}$. The minimum is at ${(0,0)}$.
 For ${q=8}$, we need to solve the cubic equation numerically, the minimum of the lowest band is
 found to be at ${(0,0)}$.

 In general, there are 5 cases in Kagome lattice: (1)
 $ q = n $ is odd, ${{-\pi\over q}\leq k_{x} \leq {\pi\over q}}$, there are ${q}$ minina in the spectrum
 at ${(0,{2\pi\over q}l)}, l=0,1,\cdots, q-1 $, $ q=3 $ case is shown in Fig.14(a). We find that the bandwidth
 does not satisfy the exponential law as shown in Fig.15a.
 (2) $q=2n$ with $ n $ odd,  ${{-2\pi\over q}\leq k_{x} \leq {2\pi\over q}}$,
  there are ${q\over2}$ minima at ${(0,{4\pi\over q}l)}, l=0,1,\cdots, q/2-1 $, $ q=6 $ case is shown in Fig.14(b).
   But when ${q=2}$, as shown above, the lowest energy band is completely
   flat. From Fig.15b, we can clearly see two separate straight lines.
   We divide the data into separate sets.
   For set 1 in Fig.15b1, the bandwidth falls as ${0.33e^{-0.20q}}$.
   For set 2 in Fig.15b2, we have the bandwidth falls as
   ${0.10e^{-0.20q}}$.(3)
   ${q=4n}$ with $ n $ odd, ${{-4\pi\over q}\leq k_{x} \leq {4\pi\over q}}$, there are ${q\over 4}$ minima
   at ${(0,{8\pi\over q}l),l=0,\cdot\cdot\cdot,{q/4-1}}$, $ q=12 $ is shown in Fig.14(c).
   The bandwidth falls as ${2.46e^{-0.21q}}$ as shwon in Fig.15(c).
   (4) ${q=8n}$, ${{-8\pi\over q}\leq k_{x} \leq {8\pi\over q}}, l=0,1,\cdots,q/8-1 $, there are also two subcases
   (4a) when n is even, there are ${2n}$ minima at  ${(-{\alpha\pi\over 3n},-\alpha\pi+{\alpha\pi\over 3n}+{2\pi\over n}l)}$
        ${\alpha=\pm}$. $ q=16 $ is shown in Fig.14(e).
   (4b) When n is odd, there are ${n}$ minima at ${(0,{16\pi\over q}l)}$, $
    q=24 $ is shown in Fig.14(d). For both  cases,  the bandwidth falls as
    ${13.87e^{-0.21q}}$ as shown in Fig.15(d). There are $ 3n $
    bands in all these cases.

   It seems to us that all the five cases have the same $ c $ within numerical errors,
   but with different magnitude $ A $.

\vspace{0.25cm}
\epsfig{file=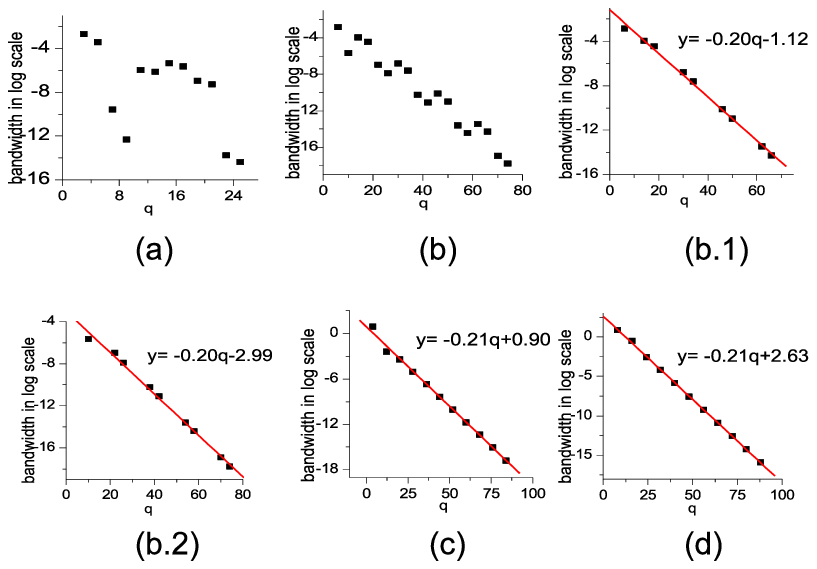,width=6in,height=4in,angle=0}
\vspace{0.25cm} {\footnotesize {\bf Fig 15:} The bandwidth of the
 bands in Kagome lattice at (a) ${q}$ is odd, (b)${q=2n}$ with $ n $ odd, (b.1)${q=2n}$ case
 1, (b.2)${q=2n}$ case 2, (c)${q=4n}$ with $ n $ odd, (d) ${q=8n}$}
\vspace{0.25cm}

\section{ Summary and conclusions}

   In this paper, we have studied the energy spectra of the Hofstadter band of vortices hopping
   on five lattices in the presence of magnetic flux $ f=p/q $ per
   smallest plaquette. Our results on  dice and Kagome  lattices are most new and interesting.
   The number of the energy bands and the number of minima in the lowest band in the five lattice
   are listed in table 1.

\begin{tabular}{|c|c|c|c|}\hline
  &square&Honeycomb& \\ \cline{2-2}\cline{3-3}
   bipartite         &$N^{n}_{q}=q $    &$N^{n}_{q}=\left\{\begin{array} {ll}
                             2 n,& n=e \\
                             n,& n=o
                              \end{array}\right.$  & \\ \cline{2-2}\cline{3-3}
            &q bands&2q bands& \\ \hline                  
            
  & triangular &dice &Kagome \\ \cline{2-2}\cline{3-3}\cline{4-4}
             
             & $N^{2n}_{q}= \left \{ \begin{array} {ll}
                             n, & n=e \\
                             2n, & n=o
                              \end{array}  \right.$,   &
            $N^{3n}_{q}= \left \{ \begin{array} {ll}
                             n, & n=e \\
                             2n, & n=o
                              \end{array}  \right.$,   & $N^{n}_{q} =n=o$ \\   \cline{2-2}\cline{3-3}\cline{4-4}
            frustrated &${q \over 2}$ bands& 3n bands& $N^{2n}_{q}=n=o$   \\   \cline{2-2}\cline{3-3}\cline{4-4}             
            & $N^{o}_{q}= q$    & $ N^{e}_{q \neq 3n}=q
    $ , & $   N^{4n}_{q}=n=o $  \\ \cline{2-2}\cline{3-3}\cline{4-4}
    &$q$ bands  & $ N^{o}_{q \neq 3n}=2 q $  & $   N^{8n}_{q}= \left \{ \begin{array} {ll}
                             2 n, & n=e \\
                             n, & n=o
                              \end{array}  \right. $   \\    \cline{3-3}\cline{4-4}
              & & above two have $3q$ bands&  all cases have $ 3n $ bands\\ \hline
                          
 \end{tabular}

\par
\vspace{0.25cm} {\footnotesize  {Table 1: The  number  of minimum
  of the lowest Hofstadter bands in the 5 lattices.  $ N^{n}_{q} $
  means $ q=n $. Suffix $ e $ and $ o $ mean even and odd. We also list the total number of bands just below each cases.
    At $ q=2 $, the lowest band in Kagome  and dice lattices is completely flat. } }

   It was argued in \cite{pq1} that for large $ q $, the bandwidth of the lowest energy Hofstadter
   band in square lattice scales as $ W= A e^{-cq} $ with $ c \sim 1 $.
   We believe that although the argument seems reasonable, it is far
   from being convincing. So it is important to test this argument
   by quantitative numerical calculations.
   We tested the rule by numerically calculating the bandwidths of the lowest Hofstadter bands
   in the 5 lattices.
   We found that this rule is indeed satisfied for all the lattices even for smallest values of $ q=1 $
   except the Kagome lattice for $ q $ is odd.  For Kogome and dice lattices, the lowest
   band is completely flat at $ q=2 $.
   The results of $ ( A, c ) $ for the five lattices are listed in the table 2:

 \vspace{0.25cm}
\begin{tabular}{ | c | c | c | c | c | } \hline
   Square &  Honeycomb  &  Triangle & Dice &  Kagome  \\  \hline
  $ ( 26.05, 1.20 ) $  &  $  ( 11.82, 1.66 ) $    & $  ( 55.70, 0.83 )_{e}  $  &
           $  ( 14.73, 0.55 )_{3n}    $      &   odd, not apply        \\  \hline
            &          &  $ ( 9.21, 0.82 )_{o} $     &  $ ( 1.01, 0.54)_{e \neq 3n}   $
                & $\begin{array} {ll}
                            &(0.33, 0.20)_{2n,1} \\
                            &(0.10, 0.20)_{2n,2}
                              \end{array}.$     \\   \hline
            &          &      &  $ ( 0.54, 0.54 )_{o \neq 3n}    $
            &$(2.46, 0.21)_{4n }$\\  \hline
            &  &  & &$(13.87, 0.21)_{8n}$\\ \hline
\end{tabular}
\par
\vspace{0.25cm} {\footnotesize  {Table 2: The  bandwidths parameters
          $ (A, c ) $ of the lowest Hofstadter bands in the 5 lattices. Suffix $ e ( o ) $
          means $ q $ is even (odd).  At $ q=2 $, the lowest band in Kagome  and dice lattices is completely flat. } }

   From table 2, it is easy to see that for a given lattice, although the
   prefactor $ A $ could be different for different cases, the $ c $
   remains the same within the numerical errors for a given lattice.
   For Kagome lattice when $ q $ is odd, the bandwidth does not satisfy any exponential
   decay law. However, in any other cases, they do satisfy the exponential laws.
   The peculiarity of Kagome lattice may be related to the fact that, in contrast to all the other 4
   lattices, Kagome lattice has both triangles which enclose $ f $
   flux quanta and hexagons which enclose $ 6f $ flux quantum.

     As said in the introduction, the first motivation to study the bandwidth is to
     look at the valid regime of dual vortex approach in the 5 lattices.
     If $ q $ is too large, the bandwidth becomes too small, the dual
     vortex approach may not be
     valid anymore.  For example, on square lattice, when $ q=4 $, $ W= 0.21 $ is already very small.
     This fact puts some doubts on the results of CDW formations in
     high temperature superconductors in \cite{pq1} where $ q $ as large as $ 8, 16,
     32 $ are used.  In fact, large $ q $
     means {\em very dilute} boson density in the direct lattice. In this
     case, the superfluid is probably the only ground state anyway
     except there are very very exotic long range interactions in
     Eqn.\ref{boson} which may stabilize CDW and VBS.
     Fortunately, the $ q=2 $ ( which is the smallest non-trivial
     case ) in honeycomb lattice was applied by one of the authors to study Helium and
     Hydrogen adsorption problems on various substrates in
     \cite{nature}.

    The second motivation is to study the tendency for interacting bosons to form a superfluid in
    the 5 lattices. The bandwidth is proportional to the vortex hopping matrix element,
    so the smallness of bandwidth
    favors the localization of the vortices, therefore enhance the tendency to form a superfluid.
     At given $ q $, the bandwidth $ W $ decreases in the order of
     Triangle, Square and Honeycomb lattice. The corresponding direct lattices
     are honeycomb, square and triangular lattices whose
     coordination numbers are 3, 4 and 6. It is known the the higher
     coordination, the easier for bosons to get the ordered superfluid state.
     The conclusions achieved in dual lattice are indeed consistent with our intuition in the direct lattice.

     As shown in the table 1, when $ q=2 $, the lowest bands in both Dice and Kagome
     lattices are flat. In dice lattice, the gap between the second flat band and the lowest flat band
     is $ \sqrt{6}  $. However, in Kagome lattice, the gap between the second dispersive band and the lowest flat band
     vanishes at $ \vec{k}=(0,0) $, so the second dispersive band can not be ignored even in the lowest energy limit.
     In dice lattice, all the three bands are flat, the vortices are completely inert,
     the interactions certainly favor the localization of the vortices.
     It indicates that for the original boson at half filling ( $ q=2 $ ) with nearest neighbor hopping on the Kagome lattice,
     there could only be a superfluid state. Slightly away from half
     filling, it was known that the superfluid state is stable against small number of vacancies or interstitials,
     we expect the superfluid state remains stable.
     This is in sharp contrast to bosons at half filling hopping on triangular lattice where
     there is a dispersion in the lowest band as shown
     in this paper. Due to the competition between
     the kinetic energy and the interactions between the vortices,
     there is a transition from a superfluid to a supersolid state as shown in \cite{gan}.
     The $ q=2 $ case at square \cite{pq1} and honeycomb \cite{nature} lattices
     were shown to have CDW  or VBS to superfluid transition in Ising or easy-plane limit. Slightly away
     from half filling, in the CDW or VBS side, it was shown in \cite{nature} that there
     must be a CDW or VBS supersolid state intervening between commensurate CDW
     or VBS to In-commensurate CDW or VBS in Ising or easy-plane
     limit. Obviously, the behaviors of bosons on Kagome lattice at or near half filling ( $ q=2
     $ ) are quite distinct from those in square, honey and
     triangular lattices.

     It is important to stress that the exactly flat bands at $ q=2 $ at
     Dice lattice are completely due to the special lattice
     structure of Dice lattice which localize the vortices. The dual vortex theory immediately leads
     to the boson superfluid state in the Kagome lattice. Of course, the
     bandwidth goes to zero at large $ q $ in any lattices. However,
     as stated in previous paragraphs, the dual vortex theory is not valid any more at sufficiently large $ q $.

     The main body of this paper only discusses the $ p=1 $ case. Taking complex
     conjugate on the Harper's equation $ H \psi( \vec{k} ) = E ( \vec{k}
     ) \psi( \vec{k} ) $ leads to $  H^{*}  \psi^{*}( \vec{k} ) = E ( \vec{k}
     ) \psi^{*}( \vec{k} ) $. Obviously, $ H^{*} $ corresponds to $
     -f $ which is equivalent to $ 1-f = 1-p/q $, so $ p=1 $ has the
     same energy spectra as $ p=q-1 $ for non-interacting vortices.
     Of course, vortex interactions will not have the periodicity $
     f \rightarrow 1+f $ anymore, so $ p=1 $ and $ p=q-1 $ in
     Eqn.\ref{boson} may not be equivalent.

     In a future publication, we are trying to understand by analytical methods similar to the ones used in \cite{thou}
     (1) why $ W= A e^{-cq} $ is satisfied at even
     smallest value of $ q $ ? Is this a unique feature of any tight binding model ? (2) For
     different cases on triangular and dice lattices and Kagome lattice for $ q $ is even listed in Table
     2, why $ c $ is the same within the numerical errors, while $ A $
     differs ? (3) What is the bandwidth rule in Kagome lattice for odd $ q $ ?
     We will also construct MSG's for dice and
     Kagome lattice to understand the energy spectra structure in Table 1.

      J. Ye thanks E. Fradkin for helpful discussions.

\appendix
 \begin{center}
    {\bf APPENDIX}
  \end{center}
 \renewcommand{\theequation}{A-\arabic{equation}}
  \setcounter{equation}{0}  In this appendix, we  simply list the Harper's equations in the
  symmetric method first used in \cite{pq1}. They may look different from those corresponding equations
  got by MBZ method used in the main text, but we show that both lead to the  same Hofstadter bands in the
  five lattices. This check ensure the correctness of the results in the main text.

 (1) Square lattice

\begin{eqnarray}
&&-e^{-ik_{y}}\psi_{l-1}-2\cos(k_{x}+2\pi fl)\psi_{l}-e^{ik_{y}}\psi_{l+1}=E\psi_{l} \nonumber\\
\end{eqnarray}
    where  $ l=0,\cdot\cdot\cdot,q-1; {-\pi\over q}\leq k_{x} \leq {\pi\over q} $

  (2) Honeycomb lattice

\begin{eqnarray}
&&-(1+e^{i(k_{x}+2\pi fl)})\psi^{a}_{l}-e^{ik_{y}}\psi^{a}_{l+1}=E\psi^{b}_{l}\nonumber\\
&&-(1+e^{-i(k_{x}+2\pi fl)})\psi^{b}_{l}-e^{-ik_{y}}\psi^{b}_{l-1}=E\psi^{a}_{l}\nonumber\\
&&
\end{eqnarray}
   where $ l=0,\cdot\cdot\cdot,q-1; -{\pi\over q} \leq k_{x} \leq {\pi\over q }$.

  (3) Triangular lattice

\begin{eqnarray}
&&-(e^{-ik_{y}}+e^{-i(k_{x}+k_{y}+2\pi f(2l-1))})\psi_{l-1}-(e^{ik_{2}}+e^{i(k_{x}+k_{y}+2\pi f(2l+1))})\psi_{l+1}\nonumber\\
&&-2\cos(k_{x}+2\pi fl)\psi_{l}=E\psi_{l}
\end{eqnarray}
   where  $ l=0,\cdot \cdot \cdot, q-1,{-\pi\over q}\leq k_{x} \leq {\pi\over
   q}$ for $ q $ odd, $ l=0,\cdot \cdot \cdot, q/2-1, {-2\pi\over q}\leq k_{x} \leq {2\pi\over
   q}$ for $ q $ even.

   (4) Dice lattice

\begin{eqnarray}
&&-(1+e^{i(k_{x}+2\pi3fl)})\psi^{a}_{l}-e^{ik_{2}}\psi^{a}_{l+1} =E\psi^{b}_{l}; \nonumber\\
&&-e^{i(k_{x}+2\pi3fl)}\psi^{a}_{l}-(e^{i(k_{y}-2\pi
f)}+e^{i(k_{x}+k_{y}+2\pi3f(l+1)-4\pi f)}) \psi^{a}_{l+1}=E\psi^{c}_{l}; \nonumber\\
&&-(1+e^{-i(k_{x}+2\pi3fl)})\psi^{b}_{l}-e^{-ik_{y}}\psi^{b}_{l-1}
-e^{-i(k_{x}+2\pi3fl)}\psi^{c}_{l}  \nonumber  \\
&& -(e^{-i(k_{y}-2\pi f)}+e^{-i(k_{x}+k_{y}+2\pi3fl-4\pi f)})
  \psi^{c}_{l-1}= E\psi^{a}_{l}
\end{eqnarray}
   where for $ {q \neq 3n}, l=0,\cdot\cdot\cdot,q-1, {-\pi\over q}\leq k_{x} \leq {\pi\over
   q}$. For ${q=3n}$, $ l=0,\cdots, q/3-1, {{-3\pi\over q}\leq k_{x} \leq {3\pi\over  q}}$
   .

   (5) Kagome lattice

\begin{eqnarray}
&&-(1+e^{i(k_{x}+2\pi8fl)})\psi^{a}_{l}-e^{-i2\pi
f}\psi^{c}_{l+1}-e^{-i(2\pi f+k_{y}-k_{x}-2\pi8fl)} \psi^{c}_{l}=E\psi^{b}_{l}; \nonumber\\
&&-\psi^{a}_{l-1}-e^{i2\pi f}\psi^{b}_{l-1}-e^{ik_{y}}\psi^{a}_{l}-e^{i(2\pi f+k_{y}-k{x}-2\pi8fl)} \psi^{b}_{l}=E\psi^{c}_{l};\nonumber\\
&&-(1+e^{-i(k_{x}+2\pi8fl)})\psi^{b}_{l}-\psi^{c}_{l+1}-e^{-ik_{y}}
\psi^{c}_{l}=E\psi^{a}_{l}
\end{eqnarray}
    where for ${q}$ is odd, $ l=0,\cdot\cdot\cdot,q-1, {{-\pi\over q}\leq k_{x}
    \leq {\pi\over q}}$, for ${q=2n}$ with $ n $ odd, $ l=0,\cdot\cdot\cdot,q/2-1, {{-2\pi\over
    q}\leq k_{x} \leq {2\pi\over
    q}}$.  For $ 4n $ with $ n $ odd, $ l=0,\cdot\cdot\cdot,q/4-1, {{-4\pi\over q}\leq k_{x} \leq {4\pi\over
    q}}$. For ${q=8n}$, $ l=0,\cdot\cdot\cdot,q/8-1, {{-8\pi\over q}\leq k_{x} \leq {8\pi\over q}}
    $.


\begin{thebibliography}{30}
\bibitem{cont} This bandwidth is much smaller than the Landau level cyclotron gap $ \sim 1/q $
    in the limit of large $ q $. In this limit, the magnetic flux
    through one plaquette gets very small, so one is approaching the
    continuum limit where the Landau levels have zero width
\end{thebibliography}
\end{document}